\documentclass[11pt]{article}

\usepackage{amsmath}

\newcommand{\N}{N\raise.7ex\hbox{\underline{$\circ $}}$\;$}

\begin{document}

\title{V.M. Red'kov\footnote{redkov@dragon.bas-net.by} \\
Rarita--Schwinger for spin  3/2 field and
 separation of the variables
in  static coordinates of de Sitter space, Schr\"{o}dinger tetrad
basis\footnote{Chapter 3 in:
 V.M. Red'kov. Tetrad formalism, spherical symmetry and Schr\"{o}dinger basis.
Publishing House "Belarusian Science", Minsk, 339 pages (2011).}
\\
{\small  Institute of Physics, National Academy of Sciences of
Belarus}
}

\date{}
\maketitle

\begin{quotation}
Rarita--Schwinger approach to description of a  massive spin 3/2
particle is investi\-gated in static coordinates of the de Sitter
space-time. The general covariant system, derived from the
relevant Lagrangian, is presented as a main wave equation and
additional constraints in the form of first order deferential and
algebraic  relations.  With the use of an extended Schr\"{o}dinger
tetrad basis and technique of Wigner D-functions  the separation
of the variable performed. 16 radial equations reduce to 8 ones
through diagonalization of $P$-inversion operator for spin 3/2
field.

\end{quotation}

The wave equation in de Sitter space-time for a particle with spin
3/2 can be presented in the form (we adhere notation used in
\cite{paper-1}; to the list of references given in \cite{paper-1} one important and much more earliest publication
must be added   \footnote{S. Deser, B. Zumino.  Consistent supergravity.  Phys. Lett. B.  1976. Vol.  62, P. 335 -- 337};
 author apologizes for that
inexactness)
$$
\left [\; i \gamma ^{\beta }(x)\; ( \nabla _{\beta } \; + \; \Gamma
_{\beta }(x) ) \; - \; m  \; \right ]\; \Psi _{\alpha }(x)  = 0 \; ,
$$
$$
\gamma ^{\alpha }(x)\; \Psi _{\alpha }(x)  = 0 \; , \;\; ( \nabla
_{\alpha }  \; + \; \Gamma _{\alpha }(x) )\; \Psi ^{\alpha }(x)  =
0 \; . \eqno(1)
$$

\noindent Here the wave function  $\Psi _{\alpha }(x)$ is a bispinor with respect to tetrad local rotations
and  a general covariant vector with respect to general coordinate transformations
It will be convenient to use a wave function with tetrad vector index
$$
\Psi _{l}(x) = e_{(l)} ^{\beta} \; \Psi _{\beta }(x) \;  , \;\;\;
\Psi _{\beta }(x) = e _{\beta }^{(l)} \; \Psi _{l}(x)\; . \eqno(2)
$$

\noindent Correspondingly the first equation in
 (1) takes the form
 $$
\left [ \;i \gamma ^{\alpha } \; ( \partial _{\alpha } \; + \; \Gamma
_{\alpha }(x)) \; \delta ^{l}_{k} \; +  \; e ^{\beta }_{(k)}\;
 e^{(l)}_{\beta ;\alpha } ) \; - \; m \;  \right ] \; \Psi _{l}(x) = 0 \; ;
\eqno(3)
$$

\noindent an addition to connection
$\Gamma _{\alpha }$ in  (3) can be presented as vector connection of the type by Tetrode--Weyl--Fock--Ivanenko
$$
e ^{\beta }_{(k)}\; e^{(l)}_{\beta ;\alpha } = ( L_{\alpha })_{k}
^{\;\;l} = {1 \over 2} ( J^{ab})^{\;\;l}_{k}\;
 e^{\beta }_{(a)}\; (\nabla _{\alpha } e_{(b)\beta })\; ,
 $$
 $$
( J^{ab})^{\;\;l}_{k}  =  \delta ^{a}_{k}\; g^{bl} \;  - \; \delta
^{b}_{k} \; g^{al}  \; . \eqno(4)
$$

\noindent Allowing for  (4), eq.  (3) is written as
$$
\left [ \; i \gamma ^{\alpha }(x) \; ( \partial _{\alpha } \; + \;
B_{\alpha }(x) ) \; -  \; m \;  \right ] \; \Psi  (x)  = 0 \; , \eqno(5)
$$
$$
B_{\alpha }(x) =  \Gamma _{\alpha }(x) \otimes  I \; + \; I
\otimes L_{\alpha }(x)    \;  . \eqno(6)
$$

\noindent   When using the components  $\Psi _{l}(x)$, additional constraints  in (1)
read
$$
\gamma^{l} \; \Psi_{l}(x) = 0 \; , \eqno(7a)
$$
$$
\left [ e^{(l)\alpha } \; \partial _{\alpha } \;  + \; e^{(l)\alpha
}(x) \; + \; e^{(l)\alpha }(x) \; \Gamma _{\alpha }(x)  \right ]
\Psi _{l}(x) = 0 \; . \eqno(7b)
$$

We will consider eqs.  (6)--(7) in static coordinates

$$
dS^{2}  =  (1 - r^{2} / \rho ^{2} ) c^{2} \; dt^{2} \; - \;
{dr^{2} \over  1  - r^{2}/\rho ^{2} }  \; -  \; r^{2} \; ( d\theta
^{2} \; + \; \sin ^{2} \theta  d\phi ^{2} )  \;  , \eqno(8)
$$

\noindent where  $\rho $  stands for the curvature radius. Below we will use dimensionless variables
$$
ct /\rho \; \rightarrow \; t  , \qquad   r/\rho   \; \rightarrow \; r
\; , \qquad  mc\rho /\hbar  \; \rightarrow \;  m \; . \eqno(9)
$$

\noindent Choosing a diagonal spherical tetrad
$$
x^{\alpha } = ( t, \; r, \; \theta , \; \phi ) \; , \qquad  \varphi
= 1  - r^{2} \; , \qquad  \varphi ' = d \varphi / dr \; ,
$$
$$
e^{\alpha }_{(0}  = ( \varphi ^{-1/2}, \; 0 , \;0 , \; 0 ) \; , \qquad
e^{\alpha }_{(3)} = ( 0 , \; \varphi ^{1/2}, \; 0 , \; 0 ) \; ,
$$
$$
e^{\alpha }_{(1)} = ( 0 , \; 0 , \; 1/r , \; 0 ) \; , \qquad  e^{\alpha
}_{(2)} = ( 0 , \; 0 , \; 0   , \; r^{-1} \sin^{-1} \theta ) \; .
\eqno(10)
$$

\noindent for  $\gamma ^{\alpha }(x)$  and   $B_{\alpha }(x)$ we get expressions

$$
\gamma ^{\alpha }(x)  = (\; { \gamma ^{0} \over   \sqrt{\varphi} }
\; , \; \sqrt{\varphi } \; \gamma ^{3} , \; {\gamma ^{1} \over
r} , \;
 {\gamma ^{2} \over   r \sin  \theta } \; ) \; ,
 $$
 $$
 B_{t}  = {1 \over 2} \varphi ' \;  (  j^{03} \otimes I + I \otimes J^{03} )\; ,
$$
$$
\Gamma_{r}  = 0 \; , \;\; \Gamma_{\theta }  = \sqrt{\varphi } \;
j^{31} \; , \;\; \Gamma_{\phi }  =  \sqrt{\varphi } \; \sin \theta
\; j^{32} \; + \; \cos \theta \; J^{12}  \; ,
$$
$$
L_{r}  = 0 \; , \;\; L_{\theta }  = \sqrt{\varphi } \; J^{31} \; ,
\;\; L_{\phi }  =  \sqrt{\varphi } \; \sin \theta \; J^{32} \; +
\; \cos \theta \; J^{12}  \;                . \eqno(11)
$$

Correspondingly, eq.  (5) takes the form
$$
\left \{ \; i {\gamma ^{0} \over \sqrt{\varphi } } \partial_{0} \;
+ \;
 i \sqrt{\varphi } \;  [\;  \gamma ^{3}  (\partial _{r} +{1 \over r} ) \; +  \;
{\gamma ^{1} J^{31}  \;+\;  \gamma ^{2} J^{32} \over r }
\right.
$$
$$
\left. + {\varphi '   \over 2 \varphi  } \; \gamma ^{0}\; (  j^{03}
\otimes I + I \otimes J^{03} ) \; ]  \; + \; {1 \over r}\; \Sigma
_{\theta ,\phi } \; -\; m \;  \right \} \Psi (t,r,\theta ,\phi ) =
0 \; , \eqno(12)
$$

\noindent where  $\Sigma _{\theta ,\phi }$ is determined by
$$
\Sigma _{\theta ,\phi }  = \; i \gamma ^{1} \; \partial _{\theta }
\; + \; \gamma ^{2}\; { i\partial _{\phi}\; + (I \otimes iJ^{12} +
i j^{12}  \otimes I )\cos \theta  \over \sin  \theta } \;
  \; .
\eqno(13)
$$

Further we will use  technique  extending Schr\"{o}dinger -- Pauli approach to treat
  problems with spherical symmetry (more details see in \cite{Book-2}).
  Let us search spherical solutions by diagonalizing operators ${\bf J}^{2}, J_{3}$ of the total angular momentum.
  Expressions for these can be found by gauge translation from
  usual ones in Cartesian tetrad. The latter is known from the theory in flat Minkowski space

$$
(\; i \gamma ^{a} \; \partial_{a} \; - \; m \; ) \; \Psi (x) = 0
\; , \; J _{i}^{Cart} = l_{i} \; + \; S_{i} \; ,
$$
 $$
S_{1} = i J^{23} \; , \qquad  S_{2} = i J^{31} \; , \qquad  S_{3}
= i J^{12}\; , \eqno(14)
$$

\noindent where
$$
S_{i} = {1 \over 2}\; \Sigma _{i} \otimes  I \; +  \; I \otimes
T_{i} \; , \qquad T_{i} =  \left | \begin{array}{cc} 0 & 0 \\ 0 &
\tau_{i}  \end{array} \right |   \; . \eqno(15)
$$

\noindent Wave functions for spin 3/2 particle in two tetrad gages in flat space/ cartesian and spherical,
are connected by
$$
\Psi  ^{sph} = S(\theta ,\phi )\; \Psi ^{Cart}\; ,\;\; S = \left |
\begin{array}{cc} U_{2} & 0 \\ 0 & U_{2}        \end{array} \right
| \otimes \left | \begin{array}{cc} 1 & 0 \\ 0 & O_{3}
\end{array} \right | \; , \eqno(16)
$$

\noindent where
$$
O = \left | \begin{array}{ccc}
\cos \theta \cos \phi  &  \cos \theta \sin \phi  & - \sin \theta \\
- \sin \phi  &  \cos \phi    &  0  \\
\sin \theta \cos \phi  &  \sin \theta \sin \phi & \cos \theta
\end{array} \right |, \qquad
U_{2} = \left | \begin{array}{cc}
       \cos \theta/2 \; e^{i\phi /2}   & \sin \theta/2  \; e^{-i\phi /2} \\[2mm]
       -\sin \theta /2 \; e^{i\phi /2}   & \cos \theta /2 \;  e^{-i\phi /2}
\end{array} \right |.
 $$

After calculation in accordance with
$$
J^{Cart}_{i} =   l_{i} \; +  \; S_{i}   \;  , \qquad  J_{i}  =
S(\theta ,\phi ) \; J^{Cart}_{i}\;  S^{-1}(\theta ,\phi )
$$

\noindent for components  $J_{i}$  in spherical tetrad basis  we get explicit expressions
$$
J_{1}  =  l_{1} \; + \; S_{3} {\sin \phi  \over \sin \theta  }  \;
,  \qquad  J_{2}  =  l_{2} \; + \; S_{3} {\cos \phi  \over \sin \theta
}  \; , \qquad  J_{3}  = l_{3} \; . \eqno(17)
$$

It is convenient to have  matrix $S_{3} = i
J^{12}$  as diagonal, but  we have
$$
S_{3}  = {1 \over 2} \left | \begin{array}{rrrr}
+1 &  0 & 0 & 0  \\     0 & -1 & 0 & 0  \\
0 &  0 &+1 & 0   \\     0 &  0 & 0 &-1      \end{array} \right |
 \otimes I \; + \;  I \otimes \left | \begin{array}{rrrr}
0 & 0 & 0 & 0 \\       0 & 0 &-i & 0 \\
0 &+i & 0 & 0 \\       0 & 0 & 0 & 0    \end{array} \right |  \;
\; ,
$$

\noindent we can reach this by transforming description to the so-called cyclic basis
 $\tilde{\Psi} = (I
\otimes U) \; \Psi $ in vector index

$$
\tilde{\Psi}  = U \; \Psi\; , \qquad \left | \begin{array}{c}
\tilde{\Psi}_{(0)} \\ \tilde{\Psi}_{(1)} \\ \tilde{\Psi}_{(2)} \\
\tilde{\Psi}_{(3)}
\end{array} \right |
 =
\left | \begin{array}{cccc}       1 & 0 & 0 & 0  \\
     0 & -1/\sqrt{2}  &   i/\sqrt{2}  &  0       \\
     0 & 0 & 0 & 1 \\   0 &  +1/\sqrt{2} &  i/\sqrt{2} & 0
\end{array} \right |
\left | \begin{array}{c} \Psi_{(0)} \\ \Psi_{(1)} \\ \Psi_{(2)} \\
\Psi_{(3)}
\end{array} \right | ,
$$

$$
\Psi  = U^{-1} \; \tilde{\Psi}\; , \qquad \left | \begin{array}{c}
\Psi_{(0)} \\ \Psi_{(1)} \\ \Psi_{(2)} \\ \Psi_{(3)}
\end{array} \right | =
\left | \begin{array}{cccc}
1 & 0 & 0 & 0 \\
0 & -1/\sqrt{2} & 0 & 1/\sqrt{2} \\
0 & -i/\sqrt{2} & 0 & -i/\sqrt{2} \\
0 & 0 & 1 & 0
\end{array} \right |
\left | \begin{array}{c} \tilde{\Psi}_{(0)} \\ \tilde{\Psi}_{(1)}
\\ \tilde{\Psi}_{(2)} \\ \tilde{\Psi}_{(3)}
\end{array} \right |  ;
\eqno(18a)
$$

\noindent note identities  $ U^{-1} = U^{+} \; , \;
\mbox{or} \;  (U^{-1})_{kl} =U^{*}_{lk}.$

In the result,   eq.  (5) assumes the form
$$
\left [ \; i \gamma ^{\alpha }(x) \; ( \partial _{\alpha } \; + \;
\Gamma _{\alpha }(x) \otimes  I \; + \; I \otimes
\tilde{L}_{\alpha }(x) ) \; -  \; m \; \right  ] \; \tilde{\Psi}
(x)  = 0 \; , \eqno(18b)
$$

\noindent  where
 $ \tilde{L}_{\alpha}  = U L_{\alpha} U^{-1}$. Additional constraints in cyclic basis will read
 $$
\gamma^{l} \; \Psi_{l}(x) =  \gamma^{l}  \; (U^{-1})_{lk} \;
\tilde{\Psi}_{k}(x) = U^{*}_{kl} \; \gamma^{l}  \;
\tilde{\Psi}_{k}(x) =  0 \; ,
$$
$$
\left [\; e^{(l)\alpha } \; \partial _{\alpha } \;  + \;
e^{(l)\alpha }(x) \; + \; e^{(l)\alpha }(x) \; \Gamma _{\alpha
}(x) \; \right ] \; (U^{-1})_{lk}  \Psi _{k}(x) = 0 \; .
\eqno(18c)
$$

 Transformed generators are
 $$
\tilde{J}^{ab} =  \sigma ^{ab} \otimes I \; + \; I \otimes
\tilde{j}^{ab}  \; , \qquad  \tilde{j}^{ab} = U \; j^{ab}\;
U^{-1}\; . \eqno(18d)
$$

\noindent or explicitly
$$
i\tilde{J}^{23} = {1 \over \sqrt{2}}  \left | \begin{array}{cccc}
0 & 0 & 0 & 0 \\     0 & 0 & 1 & 0 \\
0 & 1 & 0 & 1 \\     0 & 0 & 1 & 0   \end{array} \right | =
\tilde{T}^{1} \; , \qquad i\tilde{J}^{31}  = {1 \over \sqrt{2}}
\left | \begin{array}{cccc}
0 & 0 & 0 & 0 \\  0 & 0 & -i & 0 \\
0 & +i & 0 & 0 -i  \\    0 & 0 & +i & 0  \end{array} \right | =
\tilde{T}^{2} \; , \;
$$
$$
i\tilde{J}^{12}  =  \left | \begin{array}{cccc}
0 & 0 & 0 & 0 \\  0 & +1 & 0 & 0 \\
0 & 0 & 0 & 0 \\  0 & 0 & 0 & -1  \end{array} \right | =
\tilde{T}^{3} \; , \qquad i\tilde{J}^{03}  = \left |
\begin{array}{cccc}
0 & 0 & -1 & 0 \\   0 & 0 & 0 & 0 \\
-1 & 0 & 0 & 0 \\ 0 &  0  & 0 & 0   \end{array} \right | \;  .
$$

Let us turn to eq.    (12). Allowing for identities
$$
 \gamma ^{1} \sigma ^{31} \; + \;
\gamma ^{2} \sigma ^{32}    = \gamma ^{3} \; , \qquad \gamma ^{0}
\sigma ^{03}  = {1 \over 2} \;  \gamma ^{3}
$$

\noindent and separating a simplifying factor in $\tilde{\Psi }$ according to
$$
\tilde{\Psi }(x)  =  (  \epsilon ^{-i\epsilon t} /  r \varphi
^{1/4} )
  \; \tilde{\Phi }(r,\theta ,\phi ) \; ,
\eqno(19a)
$$

\noindent we arrive at the equation for   $\tilde{\Phi }(x)$

$$
\left  [ \; {\gamma ^{0} \over \sqrt{\varphi } } \epsilon  + i
\sqrt{\varphi } ( \gamma ^{3} \partial _{r}   + {\gamma ^{1}
\otimes \tilde{T}_{2} - \gamma ^{2} \otimes \tilde{T}_{1} \over r}
)   \right.
$$
$$
\left. + {\varphi ' \over 2 \varphi  }  \gamma ^{0} \tilde{J}^{03}
+ {1 \over r } \tilde{ \Sigma }_{\theta ,\phi }  -
 m \;  \right ] \; \tilde{\Phi }( r,\theta ,\phi  )  = 0 \; ,
\eqno(19b)
$$

\noindent where
$$
\tilde{\Sigma }_{\theta ,\phi }  = i \gamma ^{1} \partial _{\theta
}\;+\; \gamma ^{2}  {i\partial _{\phi}\; + \; i \tilde{S}_{3} \cos
\theta  \over \sin  \theta }    \; . \eqno(19c)
$$

\noindent Here, the matrix   $\tilde{S}_{3}$  is diagonal:
$\tilde{S}_{3} = {1 \over 2} \Sigma _{3} \otimes I\;+\; I \otimes
\tilde{T}_{3}  \;$.

There exist 16 possibilities for eigenvector of the matrix  $S_{3}$;  correspondingly we have 16
different eigenvectors for  $\vec{J} ^{2} ,\; J_{3}$:
$$
D_{-1/2} \left | \begin{array}{c} 1 \\ 0 \\ 0 \\ 0
\end{array} \right | \delta ^{0}_{l} \; , \;\;
D_{-3/2} \left | \begin{array}{c} 1 \\ 0 \\ 0 \\ 0
\end{array} \right | \delta ^{1}_{l} \; , \;\;
D_{-1/2} \left | \begin{array}{c} 1 \\ 0 \\ 0 \\ 0
\end{array} \right | \delta ^{2}_{l} \; , \;\;
D_{+1/2} \left | \begin{array}{c} 1 \\ 0 \\ 0 \\ 0
\end{array} \right | \delta ^{3}_{l} \; ,
$$

$$
D_{+1/2} \left | \begin{array}{c} 0 \\ 1 \\ 0 \\ 0
\end{array} \right | \delta ^{0}_{l} \; , \;\;
D_{-1/2} \left | \begin{array}{c} 0 \\ 1 \\ 0 \\ 0
\end{array} \right | \delta ^{1}_{l} \; , \;\;
D_{+1/2} \left | \begin{array}{c} 0 \\ 1 \\ 0 \\ 0
\end{array} \right | \delta ^{2}_{l} \; , \;\;
D_{+3/2} \left | \begin{array}{c} 0 \\ 1 \\ 0 \\ 0
\end{array} \right | \delta ^{3}_{l} \; ;
$$

\noindent remaining 8 eigenfunctions are constructed in the same manner but with other basis vectors

$$
\left | \begin{array}{c} 0 \\ 0 \\ 1 \\ 0
\end{array} \right |  \qquad \mbox{and} \qquad
\left | \begin{array}{c}  0 \\ 0 \\ 0 \\   1
\end{array} \right | \; .
$$

Thus, most general substitution for spherical wave solutions
with quantum numbers
 $j ,\; m$  can be presented as
  (bispinor is divided into two -2-spinors)

  $$
\tilde{\Phi }_{l} \left | \begin{array}{c} \tilde{\xi }_{l} \\
\tilde{\eta } _{l} \end{array} \right | \;\; , \;\; \tilde{\xi
}_{l} =    \left | \begin{array}{c} \tilde{\xi }^{1}_{l} \\
\tilde{\xi }^{2}_{l}  \end{array} \right | \;\; ,  \;\;
\tilde{\eta }_{l} =  \left | \begin{array}{c} \tilde{\eta
}_{\dot{1}}  \\  \tilde{\eta }_{\dot{2}}  \end{array}  \right  |
\; ,
$$

$$
\tilde{\xi }_{l} = \left | \begin{array}{c}
 f_{0} \; \delta ^{0}_{i}\; D_{-1/2} \; + \;
    f_{1}\; \delta ^{1}_{i} \; D_{-3/2} \;+\;
f_{2} \; \delta ^{2}_{i} \; D_{-1/2} \; + \; f_{3} \; \delta ^{3}_{i} \; D_{+1/2}   \\[2mm]
g_{0} \; \delta ^{0}_{i} \; D_{+1/2} \; + \; g_{1} \; \delta
^{1}_{i} \; D_{-1/2} \;+\; g_{2} \; \delta ^{2}_{i} \; D_{+1/2} \;
+ \; g_{3} \; \delta ^{3}_{i} \; D_{+3/2}
\end{array} \right |      \; ;
\eqno(20a)
$$

\noindent second 2-spinor $\tilde{\eta}_{i}$  has similar structure but with other radial functions

$$
      f_{i}(r) \; \Longrightarrow \;     h_{i}(r) \; ,
\; \; g_{i}(r) \; \Longrightarrow \;  \nu _{i}(r) \; . \eqno(20b)
$$

Note that for minimal $j=1/2$ we must impose restrictions
$f_{1}=0, \; h_{1}=0,   \;g_{3}=0, \;\nu_{3}=0 $.

Taking in mind  $\tilde{\Phi }_{l} = (\tilde{\xi}_{i} ,\; \tilde{\eta}_{i})$. eq.
(19) is written as

$$
\left [ \;  {\epsilon  \over \sqrt{\varphi }} \;+ \; \sqrt{\varphi
}  ( + i \sigma _{3} \partial _{r} \; + \; {\sigma _{1} \otimes
\tilde{T}_{2} - \sigma _{2} \otimes \tilde{T}_{1} \over r} + i
{\varphi ' \over 2 \varphi  } \; \tilde {j}^{03}  ) \; + \; {1
\over r } \; \tilde{\Sigma }_{\theta ,\phi } \;  \right ] \;
\tilde{\xi } = m \; \tilde{\eta } \;  , \eqno(21a)
$$

$$
\left [\; {\epsilon  \over \sqrt{\varphi }}\; +\; \sqrt{\varphi }
( - i \sigma _{3} \partial _{r} \;  + \; {\sigma _{1} \otimes
\tilde{T}_{2} - \sigma _{2} \otimes \tilde{T}_{1} \over r} + i
{\varphi ' \over 2 \varphi  }\; \tilde {j}^{03}  )\; -\; {1 \over
r } \;\tilde{\Sigma }_{\theta ,\phi } \; \right ]\; \tilde{\eta }
= m \; \tilde{\xi } \;  , \eqno(21b)
$$

\noindent where
$$
\tilde{\Sigma }_{\theta ,\phi }  = i \sigma _{1} \partial _{\theta
} \; + \;  \sigma _{2} {i\partial _{\phi }\; +\; (1/2 \sigma _{3}
\otimes I \; + \; I \otimes \tilde{T}_{3}) \cos \theta \over \sin
\theta  }   \; .
\eqno(21c)
$$

Additionally to $i\partial _{t} , \; \vec{J} ^{2}, \; J_{3}$
let us diagonalize
 an operator of spacial inversion. Its form in spherical basis can be found by gauge transformation from
 the known expression in Cartesian basis
 $$
 \Pi^{\;\;l}_{k} \; \Psi ^{Cart}_{l}( t,  - \vec{r} ) \; , \qquad
(\Pi^{\;\;l}_{k}) =   \left | \begin{array}{cccc}
0 & 0 & 0 & 0 \\  0  & -1  &   0  &   0   \\
0  &  0  &  -1  &   0   \\ 0  &  0  &   0  &  -1 \noindent
\end{array} \right |   \; , \eqno(22a)
$$

\noindent which results in
$$
(\tilde{\Pi}^{\;\;l}_{k}) =  \left| \begin{array}{cccc}
1 & 0 & 0 & 0 \\  0 & 0 & 0 & 1 \\
0 & 0 & 1 & 0 \\  0  &  1  &   0  &   0   \end{array} \right | \;
. \eqno(22b)
$$

\noindent An operator of $P$-inversion in bispinor index
will read (in spherical tetrad basis)
$$
\Pi  =      \left | \begin{array}{cccc}
0  &  0  &   0 &   -1  \\ 0  &  0  &  -1 &    0  \\
0  & -1  &   0 &    0  \\ -1  &  0  &   0 &    0   \end{array}
\right | \; . \eqno(23)
$$

\noindent From eigenvalue equation
$$
[\; (\; \Pi \otimes  \tilde{\Pi}^{\;\;l}_{k}\; ) \; \hat{P }\; ]\;
\tilde{\Phi}(r,\theta ,\phi ) = P \; \tilde{\Phi}(r,\theta ,\phi )
\eqno(24a)
$$

\noindent we find two values for parity $P$ and respective linear  restrictions  on
radial functions
$$
\nu _{0} = \delta \; f_{0} \; , \;\; \nu _{1} = \delta \; f_{3} \;
, \;\; \nu _{2} = \delta \; f_{2} \; , \;\; \nu _{3} = \delta \;
f_{1} \; ,
$$
$$
   h_{0} = \delta \; g_{0} \; , \;\; h_{1}    = \delta \; g_{3} \; , \;\;
   h_{2} = \delta \; g_{2} \; , \;\; h_{3}    = \delta \; g_{1} \; ,
\eqno(24b)
$$

\noindent where  $\delta = + 1$  corresponds to $P = (-1)^{J+1}$   and   $\delta =
-1$ is referred to $P = (-1)^{J}$.

Now we are to find radial equations. First let us consider eqs.  $(21a,b,c)$
 Using relations

$$
( \tilde{T}_{2})^{\;\;k}_{l} \; \delta ^{0}_{k} = 0 \; ,\qquad (
\tilde{T}_{2})^{\;\;k}_{l} \; \delta ^{1}_{k} = { i \over
\sqrt{2}} \; \delta ^{2}_{k} \; ,
$$
$$
(\tilde{T}_{2})^{\;\;k}_{l} \; \delta ^{2}_{k}  = { i \over
\sqrt{2}}\; ( \delta  ^{3}_{l} \;-\;\delta  ^{1}_{l}) \; , \qquad
(\tilde{T}_{2})^{\;\;k}_{l} \; \delta ^{3}_{k}  = - {i \over
\sqrt{2}} \; \delta ^{2}_{k}  \; ,
$$

\noindent we get
$$
(\sigma _{1} \otimes \tilde{T}_{2} )^{\;\;k}_{l} \; \tilde{\xi
}_{k}  = { i \over \sqrt{2}} \left | \begin{array}{c} g_{1} \;
\delta ^{2}_{l}\; D_{-1/2} \; + \; g_{2} \; (\delta^{3}_{l} \;-\;
\delta ^{1}_{l} )\; D_{+1/2} \; - \; g_{3} \; \delta ^{2}_{l}\; D_{+3/2}   \\[2mm]
f_{1} \; \delta ^{2}_{l} \; D_{-3/2} \; + \; f_{2} \; (\delta
^{3}_{l}\;-\; \delta ^{1}_{l} )\; D_{-1/2} - f_{3} \; \delta
^{2}_{l} \; D_{+1/2}
\end{array} \right | \; .
$$
$$
\eqno(25)
$$

\noindent Similarly,
allowing for
$$
(\tilde{T}_{1})^{\;\;k}_{l} \; \delta ^{0}_{k}  = 0 \; , \qquad (
\tilde{T}_{1})^{\;\;k}_{l} \delta ^{1}_{k} = {1 \over \sqrt{2}}
\delta ^{2}_{k} \; ,
$$
$$
(\tilde{T}_{1})^{\;\;k}_{l} \delta ^{2}_{k} = {i \over \sqrt{2}} (
\delta ^{3}_{l}\; + \; \delta ^{1}_{l} )\; ,  \qquad
(\tilde{T}_{1})^{\;\;k}_{l} \delta ^{3}_{k} = {1 \over
\sqrt{2}}\delta ^{2}_{k} \; ,
$$

\noindent we obtain
$$
- (\sigma _{2} \otimes \tilde{T}_{1} )^{\;\; k}_{l} \; \tilde{\xi
}_{k}  = {i \over \sqrt{2}} \left | \begin{array}{c} g_{1} \;
\delta ^{2}_{l} \; D_{-1/2} \; + \; g_{2}\; ( \delta ^{3}_{l} \; +
\;
\delta ^{1}_{l} ) \; D_{+1/2} \; + \; g_{3}\; \delta ^{2}_{l} \; D_{+3/2} \\[2mm]
-f_{1} \; \delta ^{2}_{l} \; D_{-3/2} \; -\; f_{2}\; ( \delta
^{3}_{l} \; + \; \delta ^{1}_{l} ) \; D_{-1/2} \; -\; f_{3}\;
\delta ^{2}_{l}\; D_{+1/2}
\end{array} \right | \; .
$$
$$
\eqno(26)
$$

\noindent Therefore, we derive

$$
(\sigma _{1} \otimes \tilde{T}_{2} \; - \;
 \sigma _{2} \otimes \tilde{T}_{1} )^{\;\;k}_{l}\; \tilde{\xi }_{k} =
i \sqrt{2} \left | \begin{array}{c}
+ g_{1} \; \delta ^{2}_{l}\; D_{-1/2} \; + \; g_{2} \; \delta ^{3}_{l}\;  D_{+1/2} \\[2mm]
- f_{2} \; \delta ^{1}_{l}\; D_{-1/2} \; - \; g_{3} \; \delta
^{2}_{l} \; D_{+1/2}
\end{array} \right |      \;  .
\eqno(27)
$$

 Further, using identities
 $$
(\tilde{j}^{03})^{\;k}_{l}   \; \delta ^{0}_{k} = - \delta
^{2}_{l} \; , \;\; (\tilde{j}^{03})^{\;\;k}_{l} \; \delta ^{1}_{k}
=  0 \; ,
$$
$$
(\tilde{j}^{03})^{\;\;k}_{l} \; \delta ^{2}_{k} = - \delta
^{0}_{l} \; , \;\; (\tilde{j}^{03})^{\;\;k}_{l}\delta ^{3}_{k} = 0
\; ,
$$

\noindent we get
$$
i {\varphi '\over 2\varphi }\; (\tilde{j}^{03})^{\;\;k}_{l} \;
\tilde{\xi }_{k}    = i {\varphi '\over 2\varphi } \left |
\begin{array}{c}
+ f_{0} \; \delta ^{2}_{l}\; D_{-1/2} \; + \; f_{2} \; \delta ^{0}_{l} \; D_{-1/2}  \\[2mm]
+ g_{0} \; \delta ^{2}_{l}\; D_{+1/2} \; + \; g_{2} \; \delta
^{0}_{l} \; D_{+1/2}
\end{array} \right |      \; .
\eqno{28}
$$

 To find $\tilde{\Sigma }_{\theta ,\phi }$ Ё§ $(21c)$ , one should use the known
 properties of Wigner function
\cite{Varshalovich-Moskalev-Hersonskiy-1975}:
$$
\partial _{\theta }\; D_{+1/2}  = {1 \over 2} \left ( a \; D_{-1/2} - b\; D_{+3/2} \right ) \; ,
$$
$$
\left [ (\sin ^{-1}(i\partial _{\phi } - {1 \over 2} \cos \theta )
\right  ]\; D_{+1/2} = {1\over 2}  \left (- a \; D_{-1/2} - b \;
D_{+3/2} \right )\;  ;
$$
$$
\partial _{\theta } \; D_{-1/2}  = {1 \over 2} \left ( b\; D_{-3/2} - a \; D_{+1/2} \right ) \; ,
$$
$$
\left [ (\sin ^{-1} \phi  (i\partial  +{1\over 2} \cos \theta )
\right  ]\; D _{-1/2} = {1 \over 2} \left (- b\; D _{-3/2} - a\;
D _{+1/2} \right  ) \; ;
$$
$$
\partial _{\theta } \; D_{+3/2}  = {1 \over 2} \left ( b \; D_{+1/2} - c\; D_{+5/2} \right ) \; ,
$$
$$
\left [ (\sin ^{-1}) (i\partial _{\phi }- {3 \over 2} \cos \theta
) \right ] \; D_{+3/2} = {1 \over 2} \left (- b\; D_{+1/2} - b \;
D_{+5/2} \right )\;  ;
$$
$$
\partial _{\theta } \; D_{-3/2}  = {1 \over 2} \left ( c \; D_{-5/2} - b\; D_{-1/2} \right )\;  ,
$$
$$
\left [ (\sin ^{-1}) (i\partial _{\phi }+{3\over 2}\cos \theta )
\right ]\; D_{-3/2} = {1 \over 2} \left (- c \; D_{-5/2} - b \;
D_{-1/2} \right )\; , \eqno(29a)
$$

\noindent where
$$
a = (j + 1/2) \;  , \qquad  b = \sqrt{(j - 1/2)(j + 3/2)}\; ,
$$
$$
c = \sqrt{(j - 3/2)(j + 5/2) } \; . \eqno{29b}
$$

\noindent So we get

$$
( \tilde{\Sigma }_{\theta ,\phi } \; \tilde{\xi } )_{l} = i \left
| \begin{array}{c} +g_{0} \; \delta ^{0}_{l} \; a \; D_{-1/2} \; +
\; g_{1} \; \delta ^{1}_{l} \; b \;D_{-3/2} \;+\;
  g_{2}\; \delta ^{2}_{l} \; a \; D_{-1/2} \; + \; g_{3} \; \delta ^{3}_{l} \; b\;D_{+1/2}  \\[2mm]
-f_{0} \; \delta ^{0}_{l}\; a \; D_{+1/2} \; - \; f_{1} \; \delta
^{1}_{l} \;b\;D_{-1/2}\;+\;
 f_{2} \; \delta ^{2}_{l}\; a \; D_{+1/2} \; -\; f_{3} \;\delta ^{3}_{l}\; b\;D_{+3/2}
\noindent \end{array} \right | \; . \eqno(30)
$$

\noindent For the term  $i \sigma _{3} \partial _{r}\; \tilde{\xi }_{l}$
we obtain

$$
i \sigma _{3} \partial _{r} \; \tilde{\xi }_{l}  = i \left |
\begin{array}{c} +f' _{0} \; \delta ^{0}_{l} \; D_{-1/2} \;+\;
f'_{1} \; \delta ^{1}_{l}\; D_{-3/2} \;+\;
 f' _{2} \; \delta ^{2}_{l} \; D_{-1/2} \;+\; f'_{3} \; \delta ^{3}_{l}\; D_{+1/2}  \\[2mm]
-g' _{0} \; \delta ^{0}_{l} \; D_{+1/2} \;-\; g'_{1} \; \delta
^{1}_{l}\; D_{-1/2} \;-\;
 g' _{2} \; \delta ^{2}_{l} \; D_{+1/2} \;-\; g'_{3} \; \delta ^{3}_{l}\; D_{+3/2}
\end{array} \right |.
\eqno{31}
$$

 With the use of the above relations
 we readily derive 8 radial equation

$$
{\epsilon  \over \sqrt{\varphi }} \; f_{0} \;+\;i \sqrt{\varphi
}\;{d \over dr}\; f_{0} \;-\; i { \varphi ' \over 2 \sqrt{\varphi
}}\; f_{2} \;+\;i {a \over r}\; g_{3} = m \; h_{0} \; ,
$$
$$
{\epsilon  \over \sqrt{\varphi }}\; g_{0} \;-\;i \sqrt{\varphi }\;
{d \over dr}\; g_{0} \;-\; i { \varphi ' \over 2 \sqrt{\varphi }}
\; g_{2} \;-\; i {a \over r} \; f_{0} = m\; \nu _{0} \; .
\eqno(32a)
$$

$$
{\epsilon \over \sqrt{\varphi }}\; f_{1} \;+\; i
\sqrt{\varphi}\;{d \over dr}\; f_{1}\;+\; i { b \over r }\; g_{1}
= m\; h_{1} \; ,
$$
$$
{\epsilon  \over \sqrt{\varphi }} \; g_{1} \;-\; i
\sqrt{\varphi}\; {d \over dr}\; g_{1} \; - \; i{\sqrt{2 \varphi }
\; \over r }\; f_{2} \;-\; i { b \over r} \; f_{1} = m\; \nu _{1}
\;  . \eqno(32b)
$$

$$
{\epsilon  \over \sqrt{\varphi }} \; f_{2}\;+\;i \sqrt{\varphi} \;
{d \over dr}\; f_{2}\;-\; i{\varphi ' \over 2 \sqrt{\varphi }} \;
f_{0} \;+\; i{\sqrt{2 \varphi } \over r}\; g_{1} \;+\; i {a \over
r} \; g_{2} = m \; h_{2} \; ,
$$
$$
{\epsilon \over \sqrt{\varphi }}\; g_{2} \;- \; i \sqrt{\varphi}\;
{d \over dr}\; g_{2} \;-\; i{\varphi ' \over 2 \sqrt{\varphi }} \;
g_{0} \;-\; i{\sqrt{2 \varphi } \over r}\; f_{3} \;-\; i{a \over
r}\; f_{2}  = m \; \nu _{2} \; , \eqno(32c)
$$

$$
{\epsilon \over \sqrt{\varphi }}\; f_{3} \;+\; i \sqrt{\varphi }
\;{d \over dr} \; f_{3}\;+\; i{\sqrt{2 \varphi } \over r}\; g_{2}
\;+\; i{b \over r}\; g_{3}  = m \; h_{3} \; ,
$$
$$
{\epsilon \over \sqrt{\varphi }}\; g_{3} \;-\; i \sqrt{\varphi}\;
{d \over dr} \; g_{3} \;-\; i { b\over r }\; f_{3} = m \; \nu _{3}
\; . \eqno(32d)
$$

These 8 equation can be translated to a new (angular) variable
$\omega$:
$$
r = \sin \omega \; ,  \;  \sqrt{\varphi }  = \cos \omega  \; ,  \;
\varphi '  = - 2 \sin \omega \; , \;\; \sqrt{\varphi } {d \over
dr}= {d \over d\omega} \; .
$$

\noindent which results in

$$
{\epsilon  \over \cos \omega } \; f_{0} \;+\; i{d \over d\omega}\;
f_{0}\;+\; i \; \mbox{tg}\; \omega \; f_{2} \;+\; i{a \over \sin
\omega} \; g_{0} = m\; h _{0}  \; ,
$$
$$
{\epsilon \over \cos \omega} \; g_{0} \;-\; i{d \over d\omega} \;
g_{0}\;+\; i \; \mbox{tg}\; \omega \; g_{2}\;-\; i{a \over \sin
\omega}\; f_{0} = m \; \nu _{0} \; ,
$$
$$
{\epsilon \over \cos \omega} \; f_{1} \;+\; i{d \over d\omega}\;
f_{1} \;+\; i{b \over \sin \omega}\; g_{1} = m \; h _{1}  \; ,
$$
$$
{\epsilon \over \cos \omega } \; g_{1}\;-\; i{d \over d\omega } \;
g_{1}\;-\; i {\sqrt{2}\over \mbox{tg}\;  \omega} \; f_{2} \;-\;
i{b \over \sin  \omega} \; f_{1} = m \; \nu _{1} \; ,
$$
$$
{\epsilon  \over \cos  \omega} \; f_{2} \;+\; i {d \over d\omega}
\; f_{2} \;+\; i \;\mbox{tg}\;  \omega \; f_{0}\; + \; i
{\sqrt{2}\over \mbox{tg}\;  \omega} \; g_{1} \;+\; i {a \over \sin
\omega} \; g_{2} = m \; h _{2} \; ,
$$
$$
{\epsilon  \over \cos  \omega} \; g_{2}\;-\; i {d \over d\omega
}\; g_{2} \;+\; i \; \mbox{tg}\;  \omega \; g_{0} \; - \; i
{\sqrt{2} \over \mbox{tg}\;  \omega}\; f_{2} - i{a \over \sin
\omega} \; f_{2} = m \; \nu _{2} \; ,
$$
$$
{\epsilon  \over \cos  \omega} \; f_{3} \;+\; i {d \over
d\omega}\; f_{3}\;+\; i{\sqrt{2} \over \mbox{tg}\; \omega}\; g_{2}
\;+\; i {b \over \sin  \omega} \; g_{3} = m \; h _{3} \; ,
$$
$$
{\epsilon  \over \cos \omega}\; g_{3} \;-\; i {d \over d\omega}\;
g_{3} \;-\; i { b \over \sin  \omega} \; f_{3} = m \; \nu _{3} \;
. \eqno(33)
$$

Performing formal changes according to

$$
f_{l}(r) \; \Longleftrightarrow  \;  h_{l}(r)\;  , \;\; g_{l}(r)
\; \Longleftrightarrow  \; \nu _{l}(r)
$$

\noindent and making some small manipulation with signs
(compare $(21a)$  with $(21b)$), we derive 8 remaining radial equations
$$
{\epsilon \over \cos \omega}\; \nu _{0} \;+\; i{d \over d\omega}
\;\nu _{0}\;+\; i \;\mbox{tg}\; \omega \; \nu _{2} \;+\; i{a \over
\sin \omega} \; h_{0} = m \; g _{0} \; ,
$$

$$
{\epsilon  \over \cos \omega}\; h_{0} \;-\; i{d \over d\omega}\;
h_{0}\;+\; i \mbox{tg}\; \omega \; h_{2}\;-\; i{a \over \sin
\omega}\; \nu _{0} = m \; f _{0} \; ,
$$

$$
{\epsilon  \over \cos \omega}\; \nu _{3} \;+\; i{d \over
d\omega}\; \nu _{3}\;+\; i {b \over \sin \omega} \; h_{3} = m \;
g_{3} \; ,
$$

$$
{\epsilon \over \cos \omega}\; h_{3} \;-\; i{d \over d\omega}\;
h_{3} \;-\; i{\sqrt{2}\over \mbox{tg}\; \omega} \; \nu _{2}
\;-\;i{b \over \sin \omega}\; \nu _{3} = m \;f _{3} \; ,
$$

$$
{\epsilon  \over \cos \omega}\; \nu _{2} \;+\; i {d \over
d\omega}\; \nu _{2} \;+\; i \; \mbox{tg}\; \omega\; \nu _{0} \;+\;
i {\sqrt{2} \over \mbox{tg}\; \omega} \; h_{3} \; + \; i {a \over
\sin \omega} \; h_{2} = m \; g_{2} \; ,
$$

$$
{\epsilon  \over \cos \omega}\; h_{2} \;-\; i{d \over d\omega} \;
h_{2} \;+\; i \; \mbox{tg}\;  \omega \; h_{0} \;-\; i {\sqrt{2}
\over \mbox{tg}\; \omega} \; \nu _{1} \;-\; i{a \over \sin \omega}
\; \nu _{2} = m \; f _{2} \; ,
$$

$$
{\epsilon  \over \cos \omega} \; \nu _{1} \;+\; i {d \over
d\omega} \; \nu _{1}\;+\; i {\sqrt{2}\over \mbox{tg}\;  \omega} \;
h_{2} \;+\; i {b \over \sin \omega} \; h_{1} = m \; g_{1} \; ,
$$

$$
{\epsilon \over \cos \omega}\; h_{1} \;-\; i{d \over d\omega}\;
h_{1} \;-\;
 i{b \over \sin \omega}\; \nu _{1} = m\; f _{1}  \; .
\eqno(34)
$$

 This 16-equationm system can be greatly simplified
 through diagonalization of spacial inversion operator -- see
 linear restrictions  $(24b)$.

 Thus, for parity $(-1)^{J+1}$,  we get 8 equations
 for
  $f_{i}(r)$  and  $g_{i}(r)$:

$$
{\epsilon  \over \cos \omega}\; f_{0} \;+\; i{d \over d\omega}\;
f_{0}\;+\; i \; \mbox{tg}\; \omega \; f_{2} \;+\; i{a \over \sin
\omega}\; g_{0} = m \; g _{0}\; ,
$$

$$
{\epsilon \over \cos \omega} \; g_{0} \;-\; i{d \over d\omega}\;
g_{0}\;+\; i \; \mbox{tg}\; \omega \; g_{2} \;-\; i{a \over \sin
\omega}\; f_{0} = m \; f _{0}  \; ,
$$

$$
{\epsilon  \over \cos \omega}\; f_{1} \;+\; i{d \over d\omega} \;
f_{1} \;+\; i{b \over \sin \omega} \; g_{1} = m \; g _{3} \; ,
$$

$$
{\epsilon  \over \cos  \omega} \; g_{1} \;-\; i {d \over
d\omega}\; g_{1}\;-\; i {\sqrt{2}\over \mbox{tg}\; \omega}\; f_{2}
\;-\;i{b \over \sin \omega}\; f_{1} = m \; f_{3} \; ,
$$
$$
{\epsilon \over \cos \omega}\; f_{2} \;+\; i{d \over d\omega}\;
f_{2} \;+\; i \; \mbox{tg}\; \omega \; f_{0} \;+\; i
{\sqrt{2}\over \mbox{tg}\;  \omega} \; g_{1} \;+\; i {a \over \sin
\omega} \; g_{2} = m \; g _{2} \; ,
$$

$$
{\epsilon  \over \cos  \omega}\; g_{2} \;-\; i {d \over d\omega}
\; g_{2} \;+\; i \; \mbox{tg}\; \omega \; g_{0} \;- \; i
{\sqrt{2}\over \mbox{tg}\; \omega} \; f_{3} \;-\; i {a \over \sin
\omega}\; f_{2} = m \; f_{2} \; ,
$$
$$
{\epsilon \over \cos \omega} \; f_{3} \;+\; i {d \over d\omega}
\;f_{3} \;+\; i {\sqrt{2} \over \mbox{tg}\; \omega} \; g_{2}
\;+\;i {b \over \sin \omega}\; g_{3} = m\; g _{1} \; ,
$$

$$
{\epsilon \over \cos \omega} \; g_{3} \;-\; i {d \over d\omega }\;
g_{3} \;-\; i{b \over \sin \omega} \; f_{3} = m \; f_{1} \; .
\eqno(35)
$$

For parity   (-$1)^{J}$, respective equations follow from  (35) by changing  $m$ into   $-m$.

Note that for minimal value $j=1/2$, the system derived becomes slightly simpler
 (remember on restrictions
$f_{1}=0,  \; g_{3}=0,  \; b = 0, \; a =1 $):
$$
{\epsilon  \over \cos \omega}\; f_{0} \;+\; i{d \over d\omega}\;
f_{0}\;+\; i \; \mbox{tg}\; \omega \; f_{2} \;+\; {i \over \sin
\omega}\; g_{0} = m \; g _{0}\; ,
$$

$$
{\epsilon \over \cos \omega} \; g_{0} \;-\; i{d \over d\omega}\;
g_{0}\;+\; i \; \mbox{tg}\; \omega \; g_{2} \;-\; {i \over \sin
\omega}\; f_{0} = m \; f _{0}  \; ,
$$

$$
{\epsilon  \over \cos  \omega} \; g_{1} \;-\; i {d \over
d\omega}\; g_{1}\;-\; i {\sqrt{2}\over \mbox{tg}\; \omega}\; f_{2}
= m \; f_{3} \; ,
$$

$$
{\epsilon \over \cos \omega}\; f_{2} \;+\; i{d \over d\omega}\;
f_{2} \;+\; i \; \mbox{tg}\; \omega \; f_{0} \;+\; i
{\sqrt{2}\over \mbox{tg}\;  \omega} \; g_{1} \;+\;  {i \over \sin
\omega} \; g_{2} = m \; g _{2} \; ,
$$

$$
{\epsilon  \over \cos  \omega}\; g_{2} \;-\; i {d \over d\omega}
\; g_{2} \;+\; i \; \mbox{tg}\; \omega \; g_{0} \;- \; i
{\sqrt{2}\over \mbox{tg}\; \omega} \; f_{3} \;-\;  {i \over \sin
\omega}\; f_{2} = m \; f_{2} \; ,
$$

$$
{\epsilon \over \cos \omega} \; f_{3} \;+\; i {d \over d\omega}
\;f_{3} \;+\; i {\sqrt{2} \over \mbox{tg}\; \omega} \; g_{2}
 = m\; g _{1} \; .
\eqno(36)
$$

Now we  turn to additional constraints   $(7a)$ and $(7b)$).
First consider   $(7a)$;  it can be written as   (expression for $U$  see in $(18b)$)

$$
 \tilde{\xi }_{0}(x) \;+\; \sigma _{i} \; U^{-1}_{ij}
\; \tilde{\xi }_{j}(x)    = 0 \; ,\qquad
 \tilde{\eta }_{0}(x) \;-\; \sigma _{i} \; U^{-1}_{ij} \;
\tilde{\eta }_{j}(x)   = 0 \; . \eqno{37a}
$$

\noindent From the first equation in $ (37a)$ it follows

$$
f_{0} D_{-1/2} \;+\; { 1 \over \sqrt{2}} \; ( - g_{1}\; D_{-1/2}+
g_{3}\; D_{+3/2} ) \;+
$$
$$
 f_{2} \; D_{-1/2} \;-\; { 1 \over
\sqrt{2}} \; ( + g_{1}\; D _{-1/2} + g_{3} \; D_{+3/2} )  = 0 \; ,
$$
$$
g_{0} \; D_{+1/2} \;+\; { 1 \over \sqrt{2}} \; (- f_{1} \;
D_{-3/2} + f_{3} \; D_{+1/2}) \;-
$$
$$
 g_{2} \; D_{+1/2} \;+\; { 1
\over \sqrt{2}} (+ f_{1} \; D_{-3/2} + f_{3} \; D_{+1/2} ) = 0 \;
, \eqno(37b)
$$

\noindent which results in two algebraic constraints
$$
 f_{0} \;-\; \sqrt{2} \; g_{1} \;+\; f_{2}   = 0 \; , \qquad
 g_{0} \;+\; \sqrt{2} \; f_{3} \;-\; g_{2}   = 0 \; .
\eqno(38a)
$$

\noindent In the same manner, from second equation in $(37a)$ it follows
$$
 h_{0} \;+\; \sqrt{2} \; \nu _{1} \;-\; h_{2}   = 0 \;, \qquad
 \nu _{0} \;-\; \sqrt{2} \; h_{3} \;+\; \nu _{2}  = 0 \; .
\eqno(38b)
$$

Allowing for   $(24b)$,  we see that at both values of parity,
(37)  and  (38) give the same two relationships
$$
 f_{0} \;-\; \sqrt{2} \; g_{1} \;+\; f_{2}   = 0 \;,\qquad
 g_{0} \;+\; \sqrt{2} \; f_{3} \;-\; g_{2}   = 0 \; .
\eqno(39)
$$

Note that these two relations are valid at minimal
$j=1/2$ as well.

Now consider the second additional  condition  $(7b)$;  it can be presented as
$$
(\nabla_{\alpha} + \Gamma_{\alpha}) \; ( e^{(l)\alpha} \Psi_{(l)}
) = [\; e^{(l)\alpha} \; \partial_{\alpha} \; + \; e^{(l)\alpha
}_{;\alpha }(x) \;+\; e^{(l)\alpha}(x)\; \Gamma _{\alpha }(x) ] \;
\Psi _{l}(x)  = 0\; , \eqno(40)
$$

\noindent where  $\Psi _{l}(x) = U^{-1}_{lk}\; \tilde{\Psi}_{k}$.
Allowing for
$$
U^{-1}_{lk}  \delta ^{0}_{k}  = \delta ^{0}_{l} \;  , \;
U^{-1}_{lk}  \delta ^{1}_{k}  = {1 \over \sqrt{2}} \; (- \delta
^{1}_{l} \;-\; i \delta ^{2}_{l} )\; ,
$$

$$
U^{-1}_{lk}  \delta ^{2}_{k}  = \delta ^{3}_{l} \;  ,  \;\;
U^{-1}_{lk}  \delta ^{3}_{k}  = { 1 \over \sqrt{2}} \; (+ \delta
^{1}_{l} \;-\; i \delta ^{2}_{l} ) \; ,
$$

\noindent we find 2-spinors  $ \xi _{l}(x) = U^{-1}_{lk} \;
\tilde{\xi}_{k}$  and  $\eta _{l}(x) = U^{-1}_{lk} \;
\tilde{\eta}_{k}$:
$$
(\xi _{l}) =
$$
$$
\left | \begin{array}{c} \; f_{0} \; \delta ^{0}_{l}\; D_{-1/2}
\;+\; f_{1} {1 \over \sqrt{2}}\; (-\delta ^{1}_{l} - i \delta
^{2}_{l} )\; D_{-3/2} \;+\; f_{2} \;\delta ^{3}_{l} \; D_{-1/2}
\;+\; f_{3} {1 \over \sqrt{2}}\;
(+ \delta ^{1}_{l} \;-\; i \delta ^{2}_{l} )\; D_{+1/2}         \\[2mm]
 g_{0}\; \delta ^{0}_{l}\; D_{+1/2}\; +\; g_{1} {1 \over \sqrt{2}}\;
(- \delta ^{1}_{l}\; -\; i \delta ^{2}_{l} ) \; D_{-1/2} \;+\;
g_{2} \; \delta ^{3}_{l} \; D_{+1/2} \;+\; g_{3} { 1 \over
\sqrt{2}} \; (+ \delta ^{1}_{l} \;- \;i \delta ^{2}_{l} )\;
D_{+3/2}
\end{array} \right |     \;    .
$$
$$
\eqno(41)
$$

\noindent Expression for  $(\eta _{l})$ will be similar but with other
radial functions  $h_{i}(r)$  and $\nu _{i}(r)$. Further,
 we find

$$
e^{(l)\alpha } \partial _{\alpha } \; \Psi _{l} = e^{(0)0 }\;
\partial_{0} \Psi_{0} +  e^{(3)r }  \; \partial_{r} \Psi_{3} +
e^{(1)\theta }  \; \partial_{\theta} \Psi_{1} + e^{(2)\phi }  \;
\partial_{\phi} \Psi_{2}
$$

$$
= {- i\epsilon  \over \sqrt{\varphi }}
\left | \begin{array}{c} f_{0} \; D_{-1/2} \\ g_{0} \; D_{+1/2} \\
h_{0} \; D_{-1/2} \\ \nu _{0} \; D_{+1/2}
\end{array} \right |
- {1\over  \sqrt{2} r} \partial_{\theta}        \left |
\begin{array}{c}
- f_{1}\; D_{-3/2} + f_{3} \; D_{+1/2} \\  - g_{1} \; D_{-1/2}  + g_{3} \; D_{+3/2} \\
- h_{1}\; D_{-3/2} + h_{3} \; D_{+1/2} \\ -\nu_{1} \; D_{-1/2}  +
\nu _{3} \; D_{+3/2}
\end{array} \right |
$$

$$
    - \sqrt{\varphi } \partial_{r}
\left | \begin{array}{c} f_{2} \; D_{-1/2} \\ g_{2} \; D_{+1/2} \\
h_{2} \; D_{-1/2} \\ \nu _{2} \; D_{+1/2}
 \end{array} \right |
+ { i \partial _{\phi } \over \sqrt{2} r \sin  \theta} \left |
\begin{array}{c}
+ f_{1} \; D_{-3/2} + f_{3} \; D_{+1/2} \\  + g_{1} \; D_{-1/2}  + g_{3} \; D_{+3/2} \\
+ h_{1} \; D_{-3/2} + h_{3} \; D_{+1/2} \\  + \nu _{1} \;
D_{-1/2}+ \nu _{3} \; D_{+3/2}
\end{array} \right | \; .
\eqno(42)
$$

\noindent With the help of identities
$$
e^{(0)\alpha }_{\;\;;\alpha } = 0 \; , \;\; e^{(1)\alpha
}_{\;\;;\alpha } = - {1 \over r} \; \mbox{ctg}\;  \theta  \; ,
\;\; e^{(2)\alpha }_{\;\; ;\alpha } = 0 \; ,\;\; e^{(3)\alpha
}_{\;\; ;\alpha } = - \sqrt{\varphi } ( {2 \over r} + {\varphi
'\over 2\varphi } )\; ,
$$

\noindent we get
$$
e^{(l)\alpha }_{\;\; ;\alpha } \; \Psi _{l}  =
 { \mbox{ctg}\; \theta  \over \sqrt{2} r}     \left | \begin{array}{c}
f_{1} \; D_{-3/2} - f_{3} \; D_{+1/2} \\ g_{1} \; D_{-1/2}  - g_{3} \; D_{+3/2} \\
h_{1} \; D_{-3/2} - h_{3} \; D_{+1/2} \\ \nu _{1} \; D_{-1/2}  -
\nu _{3} \; D_{+3/2}
\end{array} \right | -
\sqrt{\varphi} ( {2 \over r} + {\varphi '\over 2\varphi } ) \left
|
\begin{array}{c} f_{2} \; D_{-1/2} \\ g_{2} \; D_{+1/2} \\ h_{2}
\; D_{-1/2} \\ \nu _{2} \; D_{+1/2}
\end{array} \right |   \; .
\eqno(43)
$$

\noindent As for the term  $e^{(l)\alpha }\; \Gamma _{\alpha
}\; \Psi _{l}$; taking into account

$$
e^{(l)\alpha } \; \Gamma _{\alpha} \; \Psi_{l}  = e^{(0)0 } \;
\Gamma _{0} \; \Psi_{0} +  e^{(3)r } \; \Gamma _{r} \; \Psi_{3} +
e^{(1)\theta } \; \Gamma _{\theta} \; \Psi_{1} + e^{(2)\phi } \;
\Gamma _{\phi} \; \Psi_{2} =
$$

$$
 {\varphi '\over 2\sqrt{\varphi }} \; \sigma ^{03} \Psi _{0} \;-\;
{\sqrt{\varphi} \over r}\; (\sigma ^{31} \Psi _{1} \;+\; \sigma
^{32}\Psi _{2}) - { \mbox{ctg}\;  \theta \over r}\; \sigma ^{21}
\Psi_{2} \eqno(44a)
$$

\noindent after simple calculation we get
$$
e^{(l)\alpha } \Gamma _{\alpha}  \; \Psi_{l}
$$

$$
 ={\varphi ' \over 4\sqrt{\varphi }} \; \left | \begin{array}{r}
f_{0} \; D_{-1/2} \\ - g_{0} \; D_{+1/2} \\ - h_{0} \; D_{-1/2} \\
\nu _{0} \; D_{+1/2}
\end{array} \right | - {\sqrt{\varphi} \over \sqrt{2} r} \;
\left | \begin{array}{c} g_{1} \; D_{-1/2} \\ f_{3} \; D_{+1/2} \\
\nu _{1} \; D_{-1/2} \\ h_{3} \; D_{+1/2}
\end{array} \right |
+ { \mbox{ctg}\;  \theta  \over \sqrt{2}r} \; \left |
\begin{array}{c}
-f_{1} \; D_{-3/2}   - f_{3} \; D_{+1/2} \\ + g_{1} \; D_{-1/2} + g_{3} \; D_{+3/2} \\
- h_{1} \; D_{-3/2}  - h_{3} \; D_{+1/2} \\ + \nu _{1} \; D_{-1/2}
+ \nu _{3} \; D_{+3/2}
\end{array} \right |       .
\eqno(44b)
$$

\noindent Summing  (42),  (43), and  $(44b)$,  we obtain
for (40)

$$
{ -i \epsilon  \over \sqrt{\varphi }} \left | \begin{array}{c}
f_{0} \; D_{-1/2} \\ g_{0} \; D_{+1/2} \\ h_{0} \; D_{-1/2} \\ \nu
_{0} \; D_{+1/2}
\end{array} \right | +
 {\varphi ' \over 4\sqrt{\varphi }} \; \left | \begin{array}{r}
f_{0} \; D_{-1/2} \\ - g_{0} \; D_{+1/2} \\ - h_{0} \; D_{-1/2} \\
\nu _{0} \; D_{+1/2}
\end{array} \right |
$$

$$
- \sqrt{\varphi }
( \partial _{r}  + {2 \over r}+ {\varphi '\over 2 \varphi }  )
\left | \begin{array}{c}
f_{2} \; D_{-1/2} \\ g_{2} \; D_{+1/2} \\ h_{2} \; D_{-1/2} \\
\nu _{2} \; D_{+1/2}    \end{array} \right |  - {\sqrt{\varphi }
\over \sqrt{2}r} \left | \begin{array}{c} g_{1} \; D_{-1/2} \\
f_{3} \; D_{+1/2} \\ \nu _{1} \; D_{-1/2} \\ h_{3} \; D_{+1/2}
\end{array} \right |
$$

$$
 - {1 \over  \sqrt{2} \; r} \partial _{\theta } \left | \begin{array}{c}
- f_{1} \; D_{-3/2} + f_{3} \; D_{+1/2} \\  - g_{1} \; D_{-1/2}  + g_{3}\; D_{+3/2}  \\
- h_{1} \; D_{-3/2} + h_{3} \; D_{+1/2} \\  - \nu _{1} \; D_{-1/2}
+ \nu _{3} \; D_{+3/2}
\end{array} \right |
$$

$$
-{1 \over \sqrt{2}r} \left | \begin{array}{c} f_{1}\; {i \partial
_{\phi } + 3/2 \cos \theta  \over \sin \theta} D_{-3/2} \;+\;
f_{3}\; {i \partial _{\phi } - 1/2 \cos \theta  \over \sin \theta} D_{+1/2}  \\
g_{1}\; {i \partial _{\phi } + 1/2 \cos \theta  \over \sin \theta}
D_{-1/2}  \;+\;
g_{3}\; {i \partial _{\phi } - 3/2 \cos \theta  \over \sin \theta} D_{+3/2}  \\
h_{1}\; {i \partial _{\phi } + 3/2 \cos \theta  \over \sin \theta}
D_{-3/2} \;+\;
h_{3}\; {i \partial _{\phi } - 1/2 \cos \theta  \over \sin \theta} D_{+3/2} \\
\nu _{1}\; {i \partial _{\phi} + 1/2 \cos\theta \over \sin \theta}
D_{-1/2} \;+\; \nu _{3}\; {i \partial _{\phi }-3/2 \cos \theta
\over \sin \theta} D_{+3/2}
\end{array} \right |  =0   \; .
$$

\noindent From whence, transforming two last terms with the help of
 (29), we get

$$
D_{\beta } \; \Psi ^{\beta } =
 {-i \epsilon  \over \sqrt{\varphi }} \left | \begin{array}{c}
f_{0} \; D_{-1/2} \\ g_{0} \; D_{+1/2} \\ h_{0} \; D_{-1/2} \\ \nu
_{0} \; D_{+1/2}
\end{array} \right | +
 {\varphi ' \over 4\sqrt{\varphi }} \; \left | \begin{array}{r}
f_{0} \; D_{-1/2} \\ - g_{0} \; D_{+1/2} \\ - h_{0} \; D_{-1/2} \\
\nu _{0} \; D_{+1/2}
\end{array} \right |
$$

$$
-  \sqrt{\varphi }
( \partial _{r}  + {2 \over r}+ {\varphi '\over 2 \varphi }  )
\left | \begin{array}{c}
f_{2} \; D_{-1/2} \\ g_{2} \; D_{+1/2} \\ h_{2} \; D_{-1/2} \\
\nu _{2} \; D_{+1/2}    \end{array} \right |
$$

$$
- {\sqrt{\varphi }
\over \sqrt{2}r} \left | \begin{array}{c} g_{1} \; D_{-1/2} \\
f_{3} \; D_{+1/2} \\ \nu _{1} \; D_{-1/2} \\ h_{3} \; D_{+1/2}
\end{array} \right |
- {1 \over \sqrt{2}\; r}\; \left | \begin{array}{r}
(b \; f_{1} + a \; f_{3})\; D_{-1/2} \\
(a \; g_{1} + b \; g_{3})\; D_{-1/2} \\
(b \; h_{1} + a \; h_{3})\; D_{-1/2} \\
(a \; \nu _{1} + b \; \mu_{3})\; D_{-1/2} \end{array} \right | \;
. \eqno(45)
$$

\noindent Thus, we derive four differential relations -- transform them to the variable
 $\omega$ and take into account $(19a)$):

$$
{-i\epsilon  \over \cos  \omega } \; f_{0} \; -\;  { \mbox{tg}\;
\omega \over 2}\; f_{0} \;-\; {d \over d\omega }\; f_{2} \;-\; {1
\over \sqrt{2}\; \mbox{tg}\;  \omega} \; g _{1} \;-\; {1 \over
\sqrt{2} \sin  \omega }\; (b\; f_{1} + a \; f_{3})  =  0 \;  ,
$$

$$
{-i\epsilon  \over \cos  \omega } \; g_{0} \;+\; {\mbox{tg}\;
\omega  \over 2}\; g_{0} \;-\; {d \over d\omega } \; g_{2} \;-\;
{1 \over \sqrt{2}  \; \mbox{tg}\;  \omega}\; f _{3} \;-\; {1 \over
\sqrt{2} \sin  \omega}\; (a\; g_{1} + b \;g_{3})  = 0 \; ,
$$

$$
{-i\epsilon  \over \cos  \omega }\; h_{0} \;+\; { \mbox{tg}\;
\omega  \over 2}\; h_{0} \;-\; {d \over d\omega } \; h_{2} \;-\;
{1 \over \sqrt{2} \; \mbox{tg}\;  \omega } \; \nu  _{1} \;-\; {1
\over \sqrt{2} \sin  \omega }\; (b \; h_{1} + a \; h_{3})  = 0 \;
,
$$

$$
{-i\epsilon  \over \cos  \omega }\; \nu _{0} \;-\; { \mbox{tg}\;
\omega  \over 2} \;\nu _{0} \;-\; {d \over d\omega } \; \nu _{2}
\;-\; {1 \over \sqrt{2} \; \mbox{tg}\;  \omega } \; h_{3} \;-\; {1
\over \sqrt{2} \sin \omega }\; (a \; \nu _{1} + b \; \nu _{3}) = 0
\;  .
$$
$$
\eqno(46a)
$$

Note that two  first equations in  $(46a)$  include only  $f_{l}$  and $g_{l}$, whereas two las t
contain
$h_{l}$  and  $\nu _{l}$. When taking into account restrictions  (24), from  $(46a)$ it remain only two first relations
$$
{-i\epsilon  \over \cos  \theta } \; f_{0} \; -\;  { \mbox{tg}\;
\omega \over 2}\; f_{0} \;-\; {d \over d\omega }\; f_{2} \;-\; {1
\over \sqrt{2}\; \mbox{tg}\;  \omega} \; g _{1} \;-\; {1 \over
\sqrt{2} \sin  \omega }\; (b\; f_{1} + a \; f_{3})  =  0 \;  ,
$$

$$
{-i\epsilon  \over \cos  \theta } \; g_{0} \;+\; {\mbox{tg}\;
\omega  \over 2}\; g_{0} \;-\; {d \over d\omega } \; g_{2} \;-\;
{1 \over \sqrt{2}  \; \mbox{tg}\;  \omega}\; f _{3} \;-\; {1 \over
\sqrt{2} \sin  \omega}\; (a\; g_{1} + b \;g_{3})  = 0 \; .
$$
$$
\eqno(46b)
$$

For minimal  $j=1/2$ these equation become slightly simpler
$$
{-i\epsilon  \over \cos  \theta } \; f_{0} \; -\;  { \mbox{tg}\;
\omega \over 2}\; f_{0} \;-\; {d \over d\omega }\; f_{2} \;-\; {1
\over \sqrt{2}\; \mbox{tg}\;  \omega} \; g _{1} \;-\; {1 \over
\sqrt{2} \sin  \omega }\;  f_{3}  =  0 \;  ,
$$

$$
{-i\epsilon  \over \cos  \theta } \; g_{0} \;+\; {\mbox{tg}\;
\omega  \over 2}\; g_{0} \;-\; {d \over d\omega } \; g_{2} \;-\;
{1 \over \sqrt{2}  \; \mbox{tg}\;  \omega}\; f _{3} \;-\; {1 \over
\sqrt{2} \sin  \omega}\; g_{1}  = 0 \; . \eqno(47)
$$

 Let us collect results together

 after separation the variables
 in  the equations describing the massive particle with spin  $3/2$ in de Sitter space-time,
the problem is reduced to

1) 8  main radial equations  at the parity  $(-1)^{J+1}$
 (for other parity   $(-1)^{J}$, parameter  $m$  should be changed into ­   $-m$)

$$
{\epsilon  \over \cos \omega}\; f_{0} \;+\; i{d \over d\omega}\;
f_{0}\;+\; i \; \mbox{tg}\; \omega \; f_{2} \;+\; i{a \over \sin
\omega}\; g_{0} = m \; g _{0}\; ,
$$

$$
{\epsilon \over \cos \omega} \; g_{0} \;-\; i{d \over d\omega}\;
g_{0}\;+\; i \; \mbox{tg}\; \omega \; g_{2} \;-\; i{a \over \sin
\omega}\; f_{0} = m \; f _{0}  \; ,
$$

$$
{\epsilon  \over \cos \omega}\; f_{1} \;+\; i{d \over d\omega} \;
f_{1} \;+\; i{b \over \sin \omega} \; g_{1} = m \; g _{3} \; ,
$$

$$
{\epsilon  \over \cos  \omega} \; g_{1} \;-\; i {d \over
d\omega}\; g_{1}\;-\; i {\sqrt{2}\over \mbox{tg}\; \omega}\; f_{2}
\;-\;i{b \over \sin \omega}\; f_{1} = m \; f_{3} \; ,
$$

$$
{\epsilon \over \cos \omega}\; f_{2} \;+\; i{d \over d\omega}\;
f_{2} \;+\; i \; \mbox{tg}\; \omega \; f_{0} \;+\; i
{\sqrt{2}\over \mbox{tg}\;  \omega} \; g_{1} \;+\; i {a \over \sin
\omega} \; g_{2} = m \; g _{2} \; ,
$$

$$
{\epsilon  \over \cos  \omega}\; g_{2} \;-\; i {d \over d\omega}
\; g_{2} \;+\; i \; \mbox{tg}\; \omega \; g_{0} \;- \; i
{\sqrt{2}\over \mbox{tg}\; \omega} \; f_{3} \;-\; i {a \over \sin
\omega}\; f_{2} = m \; f_{2} \; ,
$$

$$
{\epsilon \over \cos \omega} \; f_{3} \;+\; i {d \over d\omega}
\;f_{3} \;+\; i {\sqrt{2} \over \mbox{tg}\; \omega} \; g_{2}
\;+\;i {b \over \sin \omega}\; g_{3} = m\; g _{1} \; ,
$$

$$
{\epsilon \over \cos \omega} \; g_{3} \;-\; i {d \over d\omega }\;
g_{3} \;-\; i{b \over \sin \omega} \; f_{3} = m \; f_{1} \; ;
\eqno(48a)
$$

2) 2 algebraic constraints (the same for both parities)

$$
g_{1} = {f_{2} + f_{0} \over \sqrt{2}} \; , \qquad f_{3} = {g_{2}
- g_{0} \over \sqrt{2}} \; ; \eqno(48b)
$$

and two differential constraints (the same for both parities)

$$
{-i\epsilon  \over \cos  \theta } \; f_{0} \; -\;  { \mbox{tg}\;
\omega \over 2}\; f_{0} \;-\; {d \over d\omega }\; f_{2} \;-\; {1
\over \sqrt{2}\; \mbox{tg}\;  \omega} \; g _{1} \;-\; {1 \over
\sqrt{2} \sin  \omega }\; (b\; f_{1} + a \; f_{3})  =  0 \;  ,
$$

$$
{-i\epsilon  \over \cos  \theta } \; g_{0} \;+\; {\mbox{tg}\;
\omega  \over 2}\; g_{0} \;-\; {d \over d\omega } \; g_{2} \;-\;
{1 \over \sqrt{2}  \; \mbox{tg}\;  \omega}\; f _{3} \;-\; {1 \over
\sqrt{2} \sin  \omega}\; (a\; g_{1} + b \;g_{3})  = 0 \; .
$$
$$
\eqno(48c)
$$

Note that the same technique can be used to separate the variables in any space-time with spherical
symmetry, also  in presence  of any external electromagnetic field with spherical symmetry (in this
 case additional constrains
should be extended -- see in \cite{paper-1}).

\end{document}